\documentstyle[12pt,psfig,epsfig,graphicx,psfrag]{article}
\newcommand{\be}{\begin{equation}}
\newcommand{\ee}{\end{equation}}
\newcommand{\ba}{\begin{eqnarray}}
\newcommand{\ea}{\end{eqnarray}}
\def\L5{\tilde{\Lambda}}

\def\Mf{M_{(5)}}

\def\in{{(i)}}
\def\inpu{{\left(i+1\right)}}
\def\inmu{{\left(i-1\right)}}

\def\jn{{(j)}}
\def\kn{{(k)}}

\begin{document}

\begin{flushright}
hep-th/yymmxxx
\end{flushright}
\vskip 1cm
\begin{center}
{\Large {\bf
Multigravity from a  discrete extra dimension}}\\[1cm]
C. Deffayet$^{a,c,}$\footnote{deffayet@iap.fr},
J. Mourad$^{b,c,}$\footnote{mourad@th.u-psud.fr},
\\
$^a${\it GReCO/IAP\footnote{FRE 2435 du CNRS.}
 98 bis boulevard Arago, 75014 Paris, France}\\
$^b${\it Laboratoire
de Physique
Th\'eorique\footnote
{UMR 8627 du CNRS.}, B\^at. 210
, Universit\'e Paris XI, \\ 91405 Orsay Cedex, France\\
$^c$ F\'ed\'eration de recherche APC, Universit\'e  Paris VII,\\
2 place Jussieu - 75251 Paris Cedex 05, France.}
\\
\end{center}
\vskip 0.2cm

Multigravity theories are constructed from the discretization
of the extra dimension of five dimensional gravity.
Using an ADM decomposition,
the discretization is performed while
maintaining the
four dimensional diffeomorphism invariance on each
site. We relate the Goldstone bosons used to realize
nonlinearly general covariance
in  discretized gravity to the shift fields of the higher
dimensional metric. We investigate the scalar excitations of
the resulting theory and show the absence of ghosts
and massive modes; this is due to a local symmetry
inherited from the reparametrization invariance along
the fifth dimension.

\vfill
\pagebreak

\section{Introduction}

Since the work of Pauli and Fierz \cite{Fierz:1939ix}, it
is known that
the consistency of theories with
 spin 2 fields is highly non trivial.
A manifestly covariant lagrangian formulation
relies on a symmetric rank 2
tensor which contains a priori ten degrees of
freedom. Using kinetic terms with four local invariances
results in  constraints which diminish the
number of degrees of freedom to six.
The additional scalar degree
of freedom turns
out to be a ghost whose elimination, at the linear level,
dictates the Pauli-Fierz quadratic combination.
Boulware and Deser \cite{Boulware:my} have argued that
the incorporation of self
interaction generically results in the reappearance of the
pathologic scalar degree of freedom.
The same reasons render the
consistency of multigravity theories
\cite{Isham:gm,Cutler:dv,Boulanger:2000rq,Damour:2002ws},
with many interacting metrics,
highly non trivial.
Higher dimensional Einstein gravity when compactified
to four dimensions results in a finite number of massless modes
among whom is a graviton and
an infinite tower of massive modes. The truncation
to a finite number of modes is however inconsistent
\cite{Dolan:1983aa,Reuter:1988ig}.
On the other hand a procedure has been recently proposed to
obtain from five dimensional Yang-Mills theories
a four dimensional theory which approach the five dimensional one
in the infrared and which has a finite  number of modes.
The key point in the approach is to replace the extra
 component of
the vector field by bifundamental scalars
which can be viewed as arising from the latticized Wilson line
along the fifth dimension.
The goal of the present letter is to present such a construction
for five dimensional gravity and to see to which extend the
consistency of the higher dimensional theory
descends to the discretized version.

Our starting point will be the 4+1 ADM decomposition \cite{adm}
of the five dimensional metric, which will be much more
convenient than the Kaluza-Klein splitting.
In particular we will show the analogy between the shift vector
in the gravity side and the fifth component of the gauge field in the
Yang-Mills side. The bifundamental scalars in the Yang-Mills side
can be seen as providing a mapping from fundamentals of a gauge
group on a site to the fundamentals of the gauge group on the
neighbouring site. We will show that the
 corresponding object in the gravity side
is a mapping from one four dimensional manifold located on a site
in  the
discrete extra dimension to the neighbouring one.
The map reduces in the continuum limit to the shift vector of the
ADM decomposition.

The paper is organized as follows.
The second Section is devoted to the introduction
of the (4+1) ADM decomposition of the 5D metric
and the  transformations of the various resulting fields
under the 5D diffeomorphisms. In the third Section we show
how to discretize the 5D Einstein-Hilbert action
while maintaining 4D diffeomorphism invariance on each
site. We relate the link fields \cite{Arkani-Hamed:2002sp} 
used to 
realize non linearly 4D general covariance
to the shift vector field of the ADM formalism. 
In the fourth Section we discuss the spectrum of the resulting
action. We exhibit, at the quadratic level,
an additional  local symmetry inherited from the 
reparametrization
invariance along the fifth dimension. This symmetry is crucial
to show the absence of ghosts and massive scalar modes.
Our analysis suggests to reconsider previous work 
\cite{Arkani-Hamed:2003vb}
on strong coupling effects in these theories.

\section{ADM decomposition and invariances}

\begin{figure}
\psfrag{Q}{{\normalsize $Q$}}
\psfrag{P}{{\normalsize $P(x)$}}
\psfrag{PP}{{\normalsize $P'(x)$}}
\psfrag{N}{{\normalsize ${\cal N}\delta y$}}
\psfrag{NN}{{\normalsize $N^\mu \delta y$}}
\psfrag{y}{{\normalsize $\Sigma_y$}}
\psfrag{yd}{{\normalsize $\Sigma_{y+ \delta y}$}}
\epsfig{file=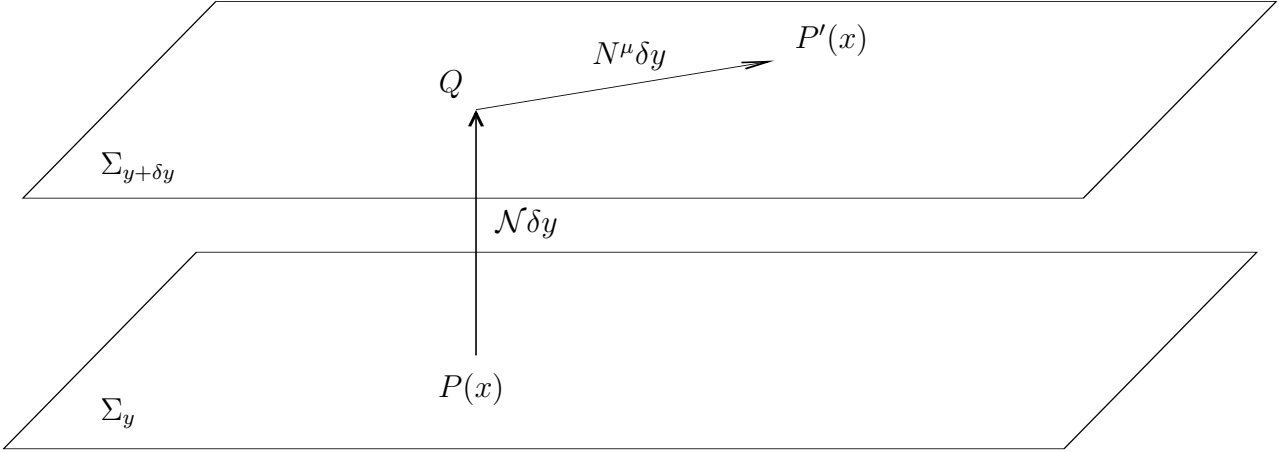, width=1.0\textwidth, height = 6cm}
\caption{The points $P$ and $P'$ have 4D coordinates $x^\mu$,
$Q$ is on the normal to $\Sigma_{y}$ at $P$.}
\label{fig2}
\end{figure}

Let us  briefly recall the ADM splitting of
the five dimensional metric $\tilde g$ and
the corresponding expression for
the Einstein-Hilbert action
\be \label{5DEH}
S_{EH}= \Mf^3 \int d^5 X  \sqrt{-\tilde{g}} \tilde{R}.
\ee
Above, and in the following, we use  expressions with a
{\it tilde} for denoting
quantities of the 5D continuum theory, like the 5D metric
$\tilde{g}_{AB}$,  upper case Latin letters from the beginning of
the alphabet, $A,B,C...$ to denote 5D indices and
lower case Greek letters, $\mu,
\nu, \alpha, ...$ will be denoting 4D indices.
We consider a  foliation of the five dimensional manifold
by four dimensional ones, $\Sigma_y$, located at given $y$.
Each slice has its four dimensional metric $g_{\mu\nu}(x,y)$.
The distance between the  manifolds $\Sigma_y$ and
$\Sigma_{y+\delta y}$ is denoted by ${\cal N} \delta y$ defining
the lapse field $\cal N$.  The normal to
$\Sigma_{y}$ at a point with 4D coordinates $x^\mu$
hits $\Sigma_{y+\delta y}$ at a point with coordinates which in
general differ from $x^\mu$ by a vector $N^\mu\delta y$ which defines
the shift field (see figure \ref{fig2}). In brief, the
 fields ${\cal N}$, ${ N}_\mu$  and $g_{\mu \nu}$
are  related to components of the 5D metric $\tilde{g}_{AB}$
by
\ba
\tilde{g}_{\mu \nu} &=& g_{\mu \nu}, \\
\tilde{g}_{\mu y} &=& N_\mu \equiv g_{\mu \alpha} N^\alpha,\\
\tilde{g}_{yy} &=& {\cal N}^2 + g_{\mu \nu} N^\mu N^\nu.
\ea
After an integration by part, the 5D Einstein-Hilbert action
(\ref{5DEH}) can be written as
\be  \label{ADM} \Mf^3
\int d^4x dy \sqrt{-g} {\cal N}
\left\{R + K_{\mu \nu} K_{\alpha \beta}
\left( g^{\mu \nu} g^{\alpha \beta} - g^{\mu \alpha}
g^{\nu \beta}\right) \right\},
\ee
where $K_{\mu \nu}$ is the extrinsic curvature of
surfaces ${\Sigma}_y$:
\be
K_{\mu \nu} = \frac{1}{2 {\cal N}}
\left(g'_{\mu \nu} - D_\mu N_\nu - D_\nu N_\mu \right),
\ee
where $D_\mu$
is the covariant derivative associated with the induced
metric $g_{\mu \nu}$ and a prime denotes an ordinary
derivative with
respect to $y$.
The equations of motions derived from action (\ref{ADM}) varying
with respect to ${\cal N}$, $N^\mu$ and $g^{\mu \nu}$ yield respectively
\ba
R &=& K^2 - K^\rho_\sigma K^\sigma_\rho \label{EQMON}\\
D^\mu K &=& D^\nu K^\mu_\nu \label{EQMONN}\\ \label{EQMOG}
G_{\mu \nu} &=& \frac{1}{2}g_{\mu \nu} \left(K^2 - K^\rho_\sigma K^\sigma_\rho\right)  + \frac{D_\mu D_\nu {\cal N} - g_{\mu \nu} D^\rho D_\rho {\cal N}}{\cal N}
- g_{\rho \mu} g_{\sigma \nu} \frac{\partial_y \left\{ \sqrt{-g} \left( K g^{\rho \sigma} - K^{\rho \sigma}\right) \right\}}{{\cal N} \sqrt{-g}} \nonumber \\
&&- \frac{2}{N}\left\{D_\nu \left(N_\mu K\right)
- D_\rho\left(K^\rho_\nu N_\mu\right) - \frac{1}{2} g_{\mu \nu}
D^\rho\left(K N_\rho \right) + \frac{1}{2} D^\rho \left(N_\rho K_{\mu \nu}\right)\right\} \nonumber \\
&&+2 \left( K^\rho_\mu K_{\rho \nu} - K K_{\mu \nu}\right),
\ea
where $G_{\mu \nu}$ is the Einstein tensor for four dimensional metric
$g_{\mu \nu}$, and $K$ is defined by $K \equiv K_{\mu \nu} g^{\mu \nu}$.

We now turn to determine the transformation properties
of the 4D fields under the various diffeomorphisms.
These are
generated by the vector fields
\be
 \tilde \xi= \xi^A\partial_A=
 \xi^\mu\partial_\mu+\xi^5\partial_y=\xi+\xi^5\partial_y,
\ee
and act on the metric as
\be
\delta \tilde g_{AB}={\cal L}_{\tilde\xi} \tilde g_{AB}=
\tilde\xi^C \partial_C \tilde g_{AB}+
\tilde g_{AC}\partial_B\tilde\xi^C
+\tilde g_{CB}\partial_A\tilde\xi^C,
\ee
where ${\cal L}_{\tilde\xi}$ is the Lie derivative with respect to $\tilde{\xi}$.
It will be very useful to write the corresponding transformations
for the 4D metric, the lapse and the shift fields.
If we define the shift vector
by
\be
\bar N=N^\mu\partial_\mu,
\ee
and the {\it covariant} derivative, $D_y$, by
\be
D_y=\partial_y-\bar N,
\ee
then, under $y$ dependent
four dimensional diffeomorphisms, that is for $\xi^5=0$,
we have
\ba
\delta g_{\mu\nu}&=&{\cal L}_{\xi}g_{\mu\nu},\label{trg}\\
\delta \bar N&=&[D_y,\xi]=\partial_y\xi-[\bar N,\xi],\label{ruln}\\
\delta {\cal N}&=&\xi({\cal N})\label{trn}.
\ea
The transformation of $g_{\mu\nu}$ and ${\cal N}$ is as expected
the one of a 4D metric and scalar respectively.
The transformation of $\bar N$ has however an
additional term with
respect
to the usual one characterizing
the transformation of a vector.
This new term is reminiscent of the inhomogeneous term
contributing to the transformation of a gauge field.
Indeed this analogy justifies the covariant derivative name
we gave to
$D_y$: suppose for example that
$\phi$ is a 5D scalar and consider $\partial_y\phi$,
it is not a scalar under a diffeomorphism generated by $\xi$
\be
\delta\partial_y\phi=\partial_y \xi(\phi)=\xi(\partial_y\phi)
+(\partial_y\xi)(\phi),
\ee
whereas $D_y\phi$ is indeed a scalar
\be
\delta D_y\phi= \delta\partial_y\phi-(\delta \bar N)(\phi)
-\bar N(\delta \phi)=
\xi(D_y\phi),
\ee
where we used the transformation rule of $\bar N$ (\ref{ruln}).
Similarly if $T$ is a tensor then ${\cal L}_{D_y} T$ is also
a tensor under 4D $y$-dependent
diffeomorphisms.
One can thus view the role of $\bar N$ as rendering possible
the formulation of an action invariant under
 $y$ dependent 4D diffeomorphisms.
Incidentally, this remark tells us how the shift fields enter
into the equations of motion: it suffices to replace everywhere
$\partial_y$ by $D_y$. Notice also that, using $D_y$,
 the extrinsic curvature
can be simply expressed as
$K_{\mu\nu}=({\cal L}_{D_y}g_{\mu\nu})/ (2{\cal N})$.

We next consider diffeomorphisms along the fifth dimension.
In fact it is more convenient to consider
diffeomorphisms generated
by $D_y$, that is $\tilde \xi=\zeta D_y$,
with $\zeta$ depending on $y$ as well as on $x$.
A short calculation gives the following rules
\ba
\delta g_{\mu\nu} &=& \zeta {\cal L}_{D_y}g_{\mu\nu},
 \ \ \delta {\cal N}=D_y(\zeta{\cal N}),\\
  \delta N^\mu &=& {\cal N}^2g^{\mu\nu}\partial_\nu \zeta.
\ea

To end this section,
let us also mention that it is also convenient to
write the action (\ref{ADM}) in the ``Einstein frame''
(from the point of view of the 4D metric $g_{\mu \nu}$).
This can be achieved
by performing the Weyl rescaling
$g_{\mu \nu} = \exp\left(-\frac{\phi}{\sqrt{3}}\right)
\gamma_{\mu \nu},$
with $ {\cal N} = \exp \left(\frac{\phi}{\sqrt{3}}\right).$
Under this transformation the action (\ref{ADM}) is rephrased into
\be  \label{ADMbis} \Mf^3
\int d^4x dy \sqrt{-\gamma} \left\{R(\gamma) -
\frac{1}{2} \gamma^{\mu \nu} \nabla_{\mu} \phi \nabla_{\nu} \phi
+ e^{-\sqrt{3} \phi}Q_{\mu \nu}
Q_{\alpha \beta}
\left(\gamma^{\mu \nu} \gamma^{\alpha \beta} -
\gamma^{\mu \alpha} \gamma^{\nu \beta} \right) \right\},
\ee
with
\ba  \label{EXTbis}
Q_{\mu \nu} &=& \frac{1}{2}
\left\{ \gamma_{\mu \nu}' - \gamma_{\mu \nu}
 \frac{\phi'}{\sqrt{3}}
- \nabla_\mu V_\nu - \nabla_\nu V_\mu  +\gamma_{\mu \nu}
V^\rho \nabla_\rho \frac{\phi}{\sqrt{3}}  \right\}\\
&=&{1 \over 2}\left[ {\cal
L}_{D_y}\gamma_{\mu\nu}-{\gamma_{\mu\nu}\over
\sqrt{3}}D_y\phi\right],
\ea
$V^\mu$ defined by $V^\mu \equiv N^\mu$,
and $V_\mu \equiv \gamma_{\mu \nu} V^\nu$.

\section{Discretization}

We first review the deconstruction of gauge theories
\cite{Arkani-Hamed:2001ca}.
Consider a 5D non-abelian gauge field
$A=A^{a}_{A}t_a dx^A\equiv
A_\mu dx^\mu +A_5d y$ with gauge group, e.g., $SO(M)$.
Under a $y$ dependent gauge transformation
the transformation rules are
\be
A'=u A u^{-1} - udu^{-1},
\ee
where $u$ is an element of $SO(M)$.
These reduce to the 4D $y$-dependent transformations for
$A_\mu dx^{\mu}$ and $A_5$, which is a scalar viewed from
4D, has the following transformation
\be
 A_5'= u A_{5} u^{-1} -u\partial_y  u^{-1},
\ee
which gives for an infinitesimal transformation
\be
\delta A_5=\partial_y\epsilon-[A_5,\epsilon].\label{tra5}
\ee
Note the formal analogy between (\ref{tra5})
and (\ref{ruln}): the Lie bracket in (\ref{ruln})
is replaced in (\ref{tra5}) by a matrix commutator.

It is very convenient when discretizing the $y$ dimension to
replace $A_5$ by the Wilson line
\cite{Wilson:1974sk,Kogut:1982ds}
\be
W(y',y)=Pe^{\int_{y}^{y'}A_5 dy},\label{wil}
\ee
which transforms as
\be
W'(y',y)=u(y')W(y',y)u^{-1}(y).
\ee
The lattice version of the gauge theory
is now easily obtained: one has $N$ sites,
on each site a gauge field with a corresponding
gauge group $SO(M)_i$
and on each link between neighbouring
sites a scalar $W(y_i,y_{i+1})$ transforming in
the bifundamental  of the gauge groups
$SO(M)_i\times SO(M)_{i+1}$.
One can then write an effective action for
the gauge fields and the scalars. The continuum limit is
recovered when the number of sites goes to infinity 
and $a$ goes to zero; the
vacuum expectation value of the scalars is then the identity:
\be
W(i,i+1)=1-a A_{5}(y)+\dots,
\ee
where $a$ is the lattice spacing.
In the broken phase one has a massless gauge boson corresponding
to
the diagonal subgroup and a collection of massive spin one
particles with masses
\be
m_k=a^{-1}\sin{k\pi \over {N}}, k=1,\dots N-1.
\ee
These reproduce, in the infrared, the Kaluza-Klein spectrum
of the first modes when the radius is given by $aN$.

We noted previously that $\bar N$ has a transformation law which
is similar
to the fifth component of the gauge potential, the
gauge group  being replaced by the diffeomorphism group.
While the gauge field can be represented by a matrix
this is no longer true
for $\bar N$ which is rather to be thought
of as an operator acting on functions on a four dimensional
manifold.
The gauge Wilson line (\ref{wil}) and
the analogy between $\bar N$
and $A_5$ motivates the consideration
of
\be
W(y',y)= P \exp \int_y^{y'} dz \bar N,\label{def1}
\ee
or more explicitly
\ba
W_{y',y} &=& 1 + \int_y^{y'} dz N^\mu(z)
\partial_\mu + \int_y^{y'} dz_1 N^{\mu_1}(z_1) \partial_{\mu_1}
\int_y^{z_1} dz_2  N^{\mu_2}(z_2) \partial_{\mu_2} + ... \nonumber \\
&& +
\int_y^{y'} dz_1 N^{\mu_1}(z_1) \partial_{\mu_1}
\int_y^{z_1} dz_2  N^{\mu_2}(z_2) \partial_{\mu_2}...
\int_y^{z_{p-1}} dz_p N^{\mu_p}(z_p) \partial_{\mu_p}
+ ... \label{DEFW}
\ea

Now $W(y',y)$ defines a mapping from functions (scalar fields)
on $\Sigma_{y}$ to functions (scalar fields)
on $\Sigma_{y'}$. Explicitly, let $\phi(x)$ be a scalar field
defined on the hypersurface $\Sigma_{y_0}$ and consider
\be
\phi_{y}=W(y,y_0)(\phi). \label{fiy}
\ee
Then $\phi_y$ verifies the equation $\partial_y \phi_y
=\bar N_y(\phi_y)$ and is subject to the boundary condition
$\phi_{y_0}=\phi$. Let $\xi(y)$ generate
a $y$-dependent 4D diffeomorphism
then from the transformation of $\bar N$ given in
(\ref{ruln}) we get
\be
\delta W(y',y)= \xi(y')W(y',y)-W(y',y)\xi(y),\label{trw}
\ee
which implies that indeed $\phi_y$ defined in (\ref{fiy})
transforms under diffeomorphisms as
$\delta \phi_y=\xi(y)(\phi_y)$ if $\phi$ transforms as
$\delta\phi=\xi(y_0)(\phi)$. A convenient and useful way of writing
(\ref{fiy}) is
\be
\phi_y=\phi \circ X(y,y_0),
\ee
where $X(y,y_0)$ is a mapping from the manifold
$\Sigma_{y}$ to $\Sigma_{y_0}$ generated by $\bar N$,
that is
\be
\partial_yX^{\mu}(y,y_0;x)=N^\mu_y(x),
\ \ X^{\mu}(y_0,y_0;x)=x^\mu,
\ee
which can be written as
\ba
X^{\mu}(y,y_0;x)&=&W(y,y_0)(x^\mu),\\
&=&x^\mu+\int_{y_0}^{y} dz N^\mu(z;x)+
\int_{y_0}^{y} dz_1
N^\nu(z_1;x)\int_{y_0}^{z_1}\partial_\nu(N^\mu(z_2;x))+\dots
\label{X}\ea
the right hand side of the first line
being understood as the action of the
$W$ on the function $x^\mu$.

It is possible to extend $W$ so that it maps tensors
of arbitrary rank on $\Sigma_y$ to tensors on
$\Sigma_{y'}$. This is done with the help of
the Lie derivative as follows
\be
W(y',y)= P \exp \int_y^{y'} dz L_{\bar N}.
\ee
It reduces to the previous expression  (\ref{def1}) when acting
on scalars.
The Leibniz rule for the Lie derivative
results in a simple action of $W$ on the direct product of tensor:
\be
W(y',y)(T_1\otimes T_2)=[W(y',y)(T_1)]\otimes [W(y',y)(T_2)],
\ee
where $T_1$ and $T_2$ are arbitrary tensors on $\Sigma_{y}$.
The commutation of the Lie derivative and the exterior derivative
when acting on forms translates also
to the simple property
\be
d[W(y',y)(\omega_{y})]=W(y',y)(d\omega_{y}),
\ee
where $\omega_{y}$ is an arbitrary form defined on $\Sigma_{y}$.
The geometric interpretation of the map $W$ is clear
from figure 1: when $y$ and $y'$ are infinitesimally close $W$ maps
the point $P$ with coordinates $x$ on $\Sigma_{y}$ to the point
$Q$ with coordinates $x^\mu+\delta y N^\mu$ on $\Sigma_{y+\delta
y}$.

We are now in a position of performing the discretization of
the Einstein-Hilbert action along the $y$ direction.
We replace $y$ by $ia$ with $i$ an integer and
$a$ the lattice spacing. The fields are thus the metric on each
site $g_{\mu\nu}^{(i)}$, the lapse fields ${\cal N}^{(i)}$
and the Wilson line $W(i,i+1)$ which, as in the gauge theory,
replaces the shift vector.
The $y$ derivative, as we noted in the
previous section,
appears in the continuum in the combination $D_y$.
The Lie derivative of a tensor field
with respect to $D_y$ can be
written as
\be
{\cal L}_{D_y}T_y=\lim_{\delta y\rightarrow 0}{ W(y,y+\delta y)
T_{y+\delta
y}-T_y \over \delta y}.
\ee
>From this we see that the
simplest discrete counterpart of the Lie
derivative along $D_y$ is
\be \label{DIFFIN}
\Delta_{\cal L} T_i={ W(i,i+1)T_{i+1}-T_i \over a}.
\ee
It is now immediate to
get the discretized Einstein-Hilbert action from
(\ref{ADM})
\be \label{DISADM}
S=\Mf^3a\sum_i\int d^4x\sqrt{-g_i}{\cal N}_i\left[R(g_i)+
{1 \over 4 {\cal N}_i^2}
(\Delta_{\cal L} g_i)_{\mu\nu}(\Delta_{\cal L} g_i)_{\alpha\beta}
\left( g^{\mu \nu}_i g^{\alpha \beta}_i - g^{\mu \alpha}_i
g^{\nu \beta}_i\right)\right]\label{actd}
\ee
The action is invariant under the product of all diffeomorphism
groups associated to the points of the lattice.
Under such a transformation generated by
$\xi_i$, the different fields transform as
\ba
\delta g_i=L_{\xi_i}g_i,\ \ \delta{\cal N}_i=\xi_i({\cal N}_i),\ \
\delta W(i,i+1)=\xi_i W(i,i+1)-W(i,i+1)\xi_{i+1}.\label{trd}
\ea
These reduce in the continuum limit to (\ref{trg}),
(\ref{trn}) and (\ref{trw}).
The explicit expression of the components of
$W(i,i+1)T_{i+1}$ can be easily written down with the help
of $X^{\mu}(i,i+1;x)$,
a mapping between the manifolds at $i$ and $i+1$
which is the discrete counterpart of
$X^\mu(y,y_0;x)$ defined in (\ref{X}). In fact, we have
\be
[W(i,i+1)T_{i+1}]_{\mu_1,\dots \mu_r}(x)=
T_{i+1}(X^{\mu}(i,i+1;x))_{\nu_1,\dots\nu_r}
\partial_{\mu_1}X^{\nu_1}\dots\partial_{\mu_r}X^{\nu_r}.
\ee
The variation of the action with respect to $W(i,i+1)$ amounts to
a variation with respect to
the mappings $X^{\mu}(i,i+1;x)$.

Notice that the action is not the most general action with the
symmetries (\ref{trd}) since it descends from a 5D theory
which had also a reparametrization invariance along the $y$
direction. This will have very important consequences as we will
show in the next section.

The same discretization procedure
can be applied to action (\ref{ADMbis})
yielding directly the action with
canonical kinetic terms for the metric
on each sites
\be  \label{DISADMbis} \sum_{i}
\Mf^3 a
\int d^4x \sqrt{-\gamma_\in}
\left\{R(\gamma_\in) -
\frac{1}{2} \gamma_\in^{\mu \nu} \nabla_{\mu} \phi_\in
\nabla_{\nu} \phi_\in
+ e^{-\sqrt{3} \phi_\in}Q^\in_{\mu \nu} Q^\in_{\alpha \beta}
\left(\gamma_\in^{\mu \nu} \gamma_\in^{\alpha \beta} -
\gamma_\in^{\mu \alpha} \gamma_\in^{\nu \beta} \right) \right\},
\ee
with
\be  \label{DISEXTbis}
Q^\in_{\mu \nu} = \frac{1}{2} \left\{\Delta_{\cal L} \gamma^\in_{\mu \nu}  - \gamma^\in_{\mu \nu} \frac{\Delta_{\cal L} \phi^\in}{\sqrt{3}}\right\}.
\ee

Before closing this section let us note that a way to realize
diffeomorphism invariance on every site has been proposed in
ref. \cite{Arkani-Hamed:2002sp} with no a priori relation to an extra dimension.
The procedure amounts to the incorporation of maps between
interacting
sites. This is the analog of our $W(i,i+1)$. Our construction
shows that this map arises naturally in the discretization
procedure and allows to identify the continuum limit of the map
with the shift vector of the ADM decomposition.
In fact in the continuum limit we have, from (\ref{X})
\be
X^{\mu}(i,i+1,x)=x^\mu+a N^{\mu}(y;x)+O(a^2).
\ee
Different approaches to the deconstruction of gravity
theories relying mainly on the local Lorentz invariance
in the Cartan moving basis formalism have
been considered in \cite{autre}.

\section{Spectrum of the action}

This section is devoted to the determination of the propagating
modes that are contained in  the action (\ref{DISADM})
at the quadratic level.
Let us first make a naive counting of the degrees of freedom
(d.o.f.) that arise from a generic  action with
the symmetries we explicitly implemented in the
action \ref{DISADM}, with a finite number, $N$, of sites.
We started with $N\times 10$  d.o.f. in the $4D$ metrics,
$N$ lapse fields and $(N-1)\times4$
d.o.f. in the mappings $W(i,i+1)$. The total number of d.o.f. is thus
$15\times N-4$. The action has local invariances
with $4N$ parameters due to the 4D diffeomorphism on the
$N$ manifolds, this reduces the number of d.o.f.
to $(15\times N-4)-(4+4)N=7N-4$.
Out of these we expect to have one graviton (2 d.o.f.) and
$N-1$ massive spin 2 particles (5N-5 d.o.f.).
The remaining degrees
of freedom are expected to be shared by a number of zero modes
(scalars and vectors), which does not depend on $N$
but depends on the boundary conditions, as well as
a number of massive scalars. The latter number depends on $N$
as $2N+c$, where $c$ is a constant which depends on the
boundary conditions but which does not depend on $N$.
These scalars are potentially pathologic, they may lead
to ghosts or tachyons. For a generic multigravity
theory, ghosts and
instabilities do indeed appear \cite{Isham:gm,Boulware:my}.
The higher dimensional theory we started with, before
discretization, does not have these pathologies.
It is thus possible that the action (\ref{DISADM})
inherited the consistency of the continuum action.
The rest of this
section is devoted to the proof that this is indeed the case at
least to the quadratic order in the fluctuations.
We will find that the massive scalar modes
decouple at the quadratic level. This is due to an extra local
symmetry which
removes $2N-2$ degrees of freedom.

In order to get the standard kinetic terms for
the metrics, let us first
 perform a Weyl rescaling on the metric
 in (\ref{DISADM}). So we define
 the metrics $\gamma_{\mu\nu}^\in$ and the scalars
 $\phi^\in$ by
  $g_{\mu \nu}^\in =
 \exp \left(-\frac{\phi_\in}{\sqrt{3}}\right)\gamma_{\mu\nu}^\in$,
 ${\cal N}^\in \equiv \exp\left(\phi^\in /\sqrt{3} \right)$.
 The next step is to expand the action around the vacuum
 \be
 \gamma_{\mu\nu}^\in=\eta_{\mu\nu}+{1 \over M_p}
 h_{\mu\nu}^\in,\ \
 \phi^\in={\varphi^\in \over M_p}, \ \
 X^\mu(i,i+1)=x^\mu+{a \over M_p}n^\mu_\in,\label{fluc}
 \ee
 and to keep the quadratic fluctuations
 in the fields. In (\ref{fluc}), $M_p$ is given by
 $M_p^2=\Mf^3 a$.
 We obtain
\ba &&\int d^4x \sum_i \frac{1}{4} \left\{  \partial_\rho h^{\mu
\nu}_\in \partial_\sigma h_\in^{\alpha \beta} \left(   \eta^{\rho
\sigma}\eta_{\mu \nu} \eta_{\alpha \beta} -\eta^{\rho
\sigma}\eta_{\mu \alpha} \eta_{\nu \beta} +
2 \delta^\sigma_{(\nu}
\eta_{\mu) \beta} \delta^\rho_\alpha
 -  \eta_{\mu \nu} \delta^\sigma_\beta
\delta^\rho_\alpha -  \eta_{\alpha \beta} \delta^\sigma_\nu
\delta^\rho_\mu \right)\right. \nonumber \\ &&\left. +
\left(\Delta \left(h_{\mu\nu}^\in-{\eta_{\mu\nu}\over \sqrt{3}}
\varphi^\in\right)-2 \partial_{(\mu} n_{\nu)}^\in
\right) \left(\Delta \left(h_{\alpha\beta}^\in-
{\eta_{\alpha\beta}\over \sqrt{3}} \varphi^\in\right)
-2\partial_{(\alpha} n_{\beta)}^\in
\right)\nonumber  \left(\eta^{\mu \nu} \eta^{\alpha \beta} -
\eta^{\mu
\alpha} \eta^{\nu \beta}\right)\right.\nonumber \\&&
 \left.
-2  \partial_\mu \varphi^\in \partial_\nu
\varphi^\in \eta^{\mu \nu}
\right\}. \label{actionl}\ea
where we have defined
the finite difference operator $\Delta$ acting on a field ${\cal
F}_\in$ as \be \Delta {\cal F}_\in = \frac{{\cal F}_\inpu -{\cal
F}_\in}{a}, \ee
two spacetime indices between
a parenthesis indicates a symmetrization on these
indices weighted by a factor two
\footnote{That is to say ${\cal F}_{(\mu \nu)} =
\frac{1}{2} \left({\cal F}_{\mu \nu} +
{\cal F}_{\nu \mu}\right)$},
we have $h^{\alpha \beta} \equiv \eta^{\alpha \mu}
\eta^{\beta \nu} h_{\mu \nu}$ and $h \equiv h^{\alpha \beta}
\eta_{\alpha \beta}$.
The equations of motion derived from this action read
\ba
0&=&\partial^\alpha \partial_\mu h_{\alpha \nu}^\in +
\partial^\alpha \partial_\nu h_{\alpha \mu}^\in +
\eta_{\mu \nu} \Box h^\in
- \Box h^\in_{\mu \nu} - \eta_{\mu \nu}
\partial^\alpha \partial^\beta h^\in_{\alpha \beta} -
\partial_\mu \partial_\nu h^\in \nonumber \\
&& +\Delta \left(
\Delta\left( h_{\alpha \beta} -
\frac{\eta_{\alpha \beta}}{\sqrt{3}}
\varphi\right) - 2 \partial_{(\alpha}
n_{\beta)}\right)_{\inmu}\left(\eta_{\mu \nu}
\eta^{\alpha \beta} -
\delta_\mu^\alpha \delta_\nu^\beta \right) \label{eind}\\
0&=& \partial_\mu\left( \Delta\left( h^\in -
\frac{4}{\sqrt{3}}\phi^\in\right)
- 2 \partial^\nu n_\nu \right) -
\partial^\nu \left(\Delta\left(h^\in_{\nu \mu} -
\frac{\eta_{\nu \mu}}{\sqrt{3}}\varphi^\in \right) -
2 \partial_{(\nu}n_{\mu)} \right) \label{max}\\
0&=&\Box \varphi^\in + \frac{\sqrt{3}}{2} \Delta
\left(\Delta \left(h - \frac{4}{\sqrt{3}} \varphi\right) -
2 \partial^\alpha n_\alpha \right)_\inmu.\label{jo}
\ea
They correspond respectively to linearization of
equations (\ref{EQMOG}), (\ref{EQMONN}) and (\ref{EQMON}).

In order to exhibit the spectrum encoded in the action
(\ref{actionl}) it is convenient to perform a discrete Fourier
transformation. To each field ${\cal F}_\in$ with
${\cal F}_{(i+N)}={\cal F}_\in$
 we define
$\hat{\cal F}_\kn$ by
\be
\hat{\cal F}_\kn=\sum_j{1 \over \sqrt{N}}{\cal F}_\jn
e^{-i2\pi jk/N}.
\ee
The action (\ref{actionl}) becomes
\ba &&\int d^4x \sum_{k} \frac{1}{4}
\left\{  \partial_\rho \hat h^{\mu
\nu}_\kn \partial_\sigma \hat h_\kn^{*\alpha \beta}
\left(   \eta^{\rho
\sigma}\eta_{\mu \nu} \eta_{\alpha \beta} -\eta^{\rho
\sigma}\eta_{\mu \alpha} \eta_{\nu \beta} +
2 \delta^\sigma_{(\nu}
\eta_{\mu) \beta} \delta^\rho_\alpha
 -  \eta_{\mu \nu} \delta^\sigma_\beta
\delta^\rho_\alpha -  \eta_{\alpha \beta} \delta^\sigma_\nu
\delta^\rho_\mu \right)\right\} \nonumber \\
&& -{1\over 2}
\sum_k  \partial_\mu \hat \varphi^\kn \partial_\nu
\hat \varphi^{*\kn} \eta^{\mu \nu}-{1\over 4}
(\partial_{\mu}\hat n_{\nu}^{(0)}-
\partial_{\nu}\hat n_{\mu}^{(0)})
(\partial^{\mu}\hat n^{\nu(0)}-
\partial^{\nu}\hat n^\mu_{(0)})+\sum_{k\neq 0}
{1 \over a^2}\sin^2{\pi k\over N}\left\{
\right. \nonumber \\
&&\left.
\left(\left(\hat h_{\mu\nu}^\kn-{\eta_{\mu\nu}\over \sqrt{3}}
\hat \varphi^\kn\right)-
{2a\partial_{(\mu} \hat n_{\nu)}^\kn\over e^{i2\pi k/N}-1}
 \right)
\left( \left(\hat h_{\alpha\beta}^{*\kn}-
{\eta_{\alpha\beta}\over \sqrt{3}} \hat\varphi^{*\kn}\right)
-{2a\partial_{(\alpha} \hat n_{\beta)}^{*\kn}\over e^{-i2\pi k/N}-1}
\right)\nonumber
\left(\eta^{\mu \nu} \eta^{\alpha \beta} -
\eta^{\mu
\alpha} \eta^{\nu \beta}\right)
\right\}. \label{actionf}
\ea
The spin two and one content of the action
is easily read from the action. We have one massless spin 2
particle given by $\hat h_{\mu\nu}^{(0)}$,
one massless spin 1 particle $\hat n_\mu^{(0)}$,
one massless scalar $\hat \phi^{(0)}$
and a tower of massive spin two particles
with a spectrum given by
\be
m^2_k={1 \over a^2}\sin^2{\pi k \over N}.
\ee
The action has the local invariances
\be
\delta\hat h_{\mu\nu}^\kn=
2\partial_{(\mu} \xi_{\nu)}^\kn,\ \
\delta\hat n_\mu^\kn=
{(e^{i2\pi k/N}-1)\over a} \xi_{\mu}^\kn,\label{invd}
\ee
which show that for $k\neq 0$, the $\hat n_\mu^\kn$ are
Stuckelberg fields which are absorbed by the massive spin 2 fields
and do not propagate.
The invariances (\ref{invd})
are the linearized version of the invariance under 4D
diffeomorphisms, they are expected by construction.
Less expected is the invariance under the local
transformations
\be
\delta \hat h_{\mu\nu}^\kn=\eta_{\mu\nu}f^\kn,\
\delta \hat \varphi^\kn=\sqrt{3}f^\kn,\
\delta \hat n_\mu^\kn={a \over 1-e^{-i2\pi k/N}}\partial_\mu f ^\kn,\
k\neq 0.\label{invr}
\ee
A generic multigravity theory with
4D diffeomorphism invariance on each site realized does not
possess this symmetry, which is inherited from the
diffeomorphism invariance under the $y$ reparametrizations in the
continuum theory. In fact
the invariance under (\ref{invr}) eliminates,
at the quadratic level, all except the massless,
scalar modes $\hat\phi^\kn$. It may also be used
to eliminate the trace of the
$\hat h_{\mu\nu}^\kn$ proving the
absence of ghostlike excitations. Associated to this
local invariance there is a constraint
which removes one more set of scalars.
At this point we note that while  the Pauli-Fierz action
removes the ghost by hand, the above action removes it with the
aid of a local symmetry; in the gauge where
$\hat \varphi^\kn$ and $\hat n_\mu^\kn$ are zero the
action reduces to the
Pauli-Fierz form.

The scalar fluctuations  are in fact potentially pathologic.
This is mainly due to the fact that the conformal factor
has a kinetic term with the wrong sign, that is
 for a metric of the form $g=e^{2\sigma}\eta$, the Einstein
action reads
\be
6\int(\partial e^\sigma)^2,\label{conf}
\ee
giving a ghostlike kinetic term for $e^{\sigma}$

In order to explicitly check the consistency of the action
(\ref{actionl})
 it is instructive
to isolate  in (\ref{actionl}) the scalar massive modes
which in addition to $\hat{\varphi}^\kn$ are  given by
\ba \label{gamsca}
\hat \gamma^\kn_{\mu \nu} &=& 
\eta_{\mu \nu} \hat \psi^\kn + 
\frac{\partial_\mu \partial_\nu}{\Box} \hat f^\kn, k\neq 0 \\
\hat n_\mu^\kn &=& \partial_\mu \hat v^\kn
\ea
At quadratic order in $\hat\psi^\kn, \hat f^\kn, \hat v^\kn$
and $\hat\phi^\kn$, the action
(\ref{DISADM}) now reads
\ba \label{QUADAC}
\int d^4x\sum_k \left\{\frac{3}{2} 
\left|\partial \hat\psi^\kn\right|^2 - 
\frac{1}{2} \left|\partial \hat\varphi^\kn\right|^2  +
{12 \sin^2{\pi k\over N} \over a^2}
 \left|\left(\hat\psi^\kn - \frac{\hat\varphi^\kn}{\sqrt{3}}
\right)\right|^2\right.\nonumber\\
 \left.
+ \left[\frac{3}{a^2}\sin^2{\pi k\over N}
\left(\hat f^\kn - 
{2a\Box \hat  v^\kn \over e^{i2\pi k/N}-1}\right)
\left(\hat\psi^\kn - \frac{\hat\varphi^\kn}{\sqrt{3}}  \right)^*
 +c.c.\right]\right\}
\ea
On the above action one clearly sees the ghost like
nature of $\hat\psi^\kn$
as well as the fact that the discretization procedure is
leading to tachyonic mass term for $\hat\phi^\kn$.
We saw however that both problems  disappear once
one carefully considers the invariances of
the action (\ref{QUADAC}) given in (\ref{invd}).
The equations of motion derived from the above
action (\ref{QUADAC}) now read
\ba \label{EQPP}
\Box \hat\psi_\kn  &=& -{8\sin^2{\pi k\over N}\over a^2}
  \left(\hat \psi_\kn -
 \frac{\hat\varphi_\kn}{\sqrt{3}}\right) - 
 \frac{2\sin^2{\pi k\over N}}{a^2}
 \left(\hat f_\kn - 
 {2a\Box \hat  v^\kn \over e^{i2\pi k/N}-1}\right), \\
 \hat\psi_\kn - 
\frac{\hat\varphi_\kn}{\sqrt{3}}  &=& 0,\label{cons1}\\
\Box \hat{\varphi_\kn\over \sqrt{3}}  &=& -
{8\sin^2{\pi k\over N}\over a^2}
  \left(\hat \psi_\kn -
 \frac{\hat\varphi_\kn}{\sqrt{3}}\right) - 
 \frac{2\sin^2{\pi k\over N}}{a^2}
 \left(\hat f_\kn - 
 {2a\Box \hat  v^\kn \over e^{i2\pi k/N}-1}\right), \label{EQPHI}\\
\partial^\mu \left(\hat\psi^\kn -
\hat \frac{\varphi^\kn}{\sqrt{3}} \right) &=& 0. \label{EQV}
\ea
One can explicitly verify that these equations
are obtained from the full equations of motion
(\ref{eind}-\ref{jo}).
The first three equations are respectively
the equation of motion of
$\hat\psi_\kn$, $\hat f_\kn$, $\hat\varphi_\kn$, 
while the last is
the equation of motion for
$\partial_\mu\hat v_\kn$ considered as a dynamical variable.
Note that had we considered (as in ref.
\cite{Arkani-Hamed:2003vb}), $\hat v_\kn$,
instead of $\partial_\mu \hat  v_\kn$, 
as the dynamical variable,
we would have obtained the equation of motion
$\Box \left(\hat \psi_\kn - { \hat\varphi_\kn\over \sqrt{3}}
 \right) =0$,
instead of the constraint (\ref{EQV}) which eliminates the modes
$\left( \hat\psi^\kn -
 \frac{\hat\varphi^\kn}{\sqrt{3}} \right)$.
This would  not have been the equation
of motion obtained by varying the initial action
(\ref{ADM}) with respect to $N^\mu$.
The constraints (\ref{cons1})
when put back in the action leads
to the cancellation of the
kinetic terms of
both $\hat\psi^\kn$ and $\hat\varphi^\kn$.
It is possible to use the gauge invariance (\ref{invr})
to set $\hat \varphi^\kn=0$ for $k\neq 0$ and then use the 4D
diffeomorphisms to eliminate also $\hat v^\kn$ and
$\hat f^{(0)}$.
The equations of motion eliminate $\hat f^\kn$ for $k\neq 0$ as
well as
$\hat \psi^\kn$ for all $k$\footnote{That $\hat \psi^{(0)}$
is eliminated can be seen from the full equations (\ref{eind})}.
The only remaining scalar is thus the
massless
$\hat \varphi^{(0)}$.

At the cubic and higher orders the symmetries
(\ref{invd}) are not expected to hold anymore,
neither an extension
of these.
 The cubic part of the action is the sum of the cubic part of the
 Einstein-Hilbert action and the additional terms given by
\ba
M_pS^{(3)}=&&\sum_i
\left(\Delta \left(h_{\mu\nu}^\in-{\eta_{\mu\nu}\over \sqrt{3}}
\varphi^\in\right)-2 \partial_{(\mu} n_{\nu)}^\in
\right) \left(\Delta \left(h_{\alpha\beta}^\in-
{\eta_{\alpha\beta}\over \sqrt{3}} \varphi^\in\right)
-2\partial_{(\alpha} n_{\beta)}^\in
\right)  \nonumber\\
&&\left\{
\left(\eta^{\mu \nu} \eta^{\alpha \beta} -
\eta^{\mu
\alpha} \eta^{\nu \beta}\right)({1\over
2}h^\in-{\varphi^\in\over\sqrt{3}})-
\left(h^{\mu \nu} \eta^{\alpha \beta} +\eta^{\mu \nu} h^{\alpha \beta}-
h^{\mu\alpha} \eta^{\nu \beta}
-\eta^{\mu\alpha} h^{\nu \beta}\right)\right\}\nonumber\\
&&+2\left(\Delta \left(h_{\mu\nu}^\in-{\eta_{\mu\nu}\over \sqrt{3}}
\varphi^\in\right)-2 \partial_{(\mu} n_{\nu)}^\in
\right)\left(\eta^{\mu \nu} \eta^{\alpha \beta} -
\eta^{\mu
\alpha} \eta^{\nu \beta}\right)\nonumber\\
&&\left(
-{\Delta(h_{\alpha\beta}\phi)^\in\over\sqrt{3}}
+{\eta_{\alpha\beta}\over 6}
\Delta\phi^2_\in
-{\eta_{\alpha\beta}\over
\sqrt{3}} n^\lambda_\in\partial_\lambda\phi^{(i+1)}
-{2\over \sqrt{3}}\phi^{(i+1)}\partial_{(\alpha} n_{\beta)}^\in
\right. \nonumber\\
&&\left.
+h_{\alpha_\lambda}\partial_\beta n^\lambda_\in
+h_{\beta_\lambda}\partial_\alpha n^\lambda_\in
+n^\lambda_\in\partial_\lambda h_{\alpha\beta}^{(i+1)}
+a\partial_\alpha n^\lambda_\in\partial_\beta
n_\lambda^\in \right).
\ea
The characteristic scale of these interactions is given by
$\sqrt{NM_p/a}=\sqrt{R}M_{(5)}^{3\over 4}a^{-5\over 4}$
when $a>M_p^{-1}$
or else by $\sqrt{N}M_p=M_{(5)}^{3\over 2}\sqrt{R}$.
Notice however, that since at the quadratic
order all the scalars (except $\hat \varphi^{(0)}$) decouple
and thus have no propagators, we have a non-standard action
which starts at a cubic or higher order level for these modes.
This deserves a further study.

\section*{Acknowledgements}
We thank E. Dudas, U. Ellwanger and S. Pokorski for
helpful discussions.

\end{document}